\documentclass[draft,aps,prd,showpacs,superscriptaddress]{revtex4} 

\def\bfl{\begin{flushleft}}
\def\efl{\end{flushleft}}
\def\bfr{\begin{flushright}}
\def\efr{\end{flushright}}
\def\bc{\begin{center}}
\def\ec{\end{center}}
\def\be{\begin{equation}}
\def\ee{\end{equation}}
\def\ba{\begin{eqnarray}}
\def\ea{\end{eqnarray}}
\def\baa#1{\begin{array}{#1}}
\def\eaa{\end{array}}
\def\bw{\begin{widetext}}
\def\ew{\end{widetext}}
\def\nn{\nonumber }
\def\lb#1{\label{#1}}

\def\text#1{\mbox{#1}}

\begin{document}

\title{Transport theory in the normal state of high-$T_c$ superconductors}

\author{
Andrew Das Arulsamy \thanks{$E-mail: sadwerdna@hotmail.com}}

\affiliation{ Condensed Matter Group, No. 22, Jln Melur 14, Tmn
Melur, Ampang, Selangor DE, Malaysia}

\affiliation{ Department of Physics, National University of
Singapore, 2 Science Drive 3, Singapore 117542, Singapore}

\date{\today}

\begin{abstract}
Transport mechanism in the normal state of high-$T_c$
superconductors is described using the well known Fermi-Dirac
statistics in which an additional restrictive constraint is
introduced so as to capture the variation of resistivity with
temperature and doping. The additional restrictive condition is
the ionization energy that will eventually determine the
properties of charge carriers' in the normal state of high-$T_c$
superconductors. The magnitude and the variation of charge
carriers concentration and resistivities (polycrystalline, c-axis
and $ab$-planes) with temperature and doping are very well
described by the ionization energy based Fermi-Dirac statistics.
However, these transport models are not appropriate for cuprates
below the characteristics ($T^*$) and critical temperatures
($T_c$), metals with free electrons and strong electron-phonon
scattering. Ionization energy is found to be an essential
parameter to accurately predict variations of resistivity's
magnitude with doping, charge carriers' concentrations, scattering
rate constants as well as the effective mass. Apart from that,
iFDS based resistivity models provide the comprehensive
information on crossover-temperature (metallic $\to$ insulating
transition temperature) in the normal state of high-$T_c$
superconductors.
\end{abstract}


\maketitle

\narrowtext


\section{Introduction}\lb{s-in}

Superconductivity has been around for almost a century since its
discovery in 1911 by Kamerlingh-Onnes~\cite{onnes1}. Subsequent
discovery on Copper-oxide (Cu-O$_2$) materials by Bednorz and
Muller~\cite{bednorz2} in 1986 literally questions the
applicability of the Nobel prize winning Bardeen-Cooper-Schrieffer
(BCS) theory~\cite{bardeen5} in cuprates. This terrible yet
exciting scenario have had led to numerous theoretical proposals
and experimental endeavours due to the fact that the transport
properties of Cu-O$_2$ based ceramics are overwhelmingly
mystifying. Apart from cuprates, superconductivity seems to occur
in almost all type of materials namely, organic Bechgaard
salts~\cite{jerom5a,bechgaard5b}, ferromagnetic~\cite{pf5c}
ZrZn$_2$, metallic Magnesium diboride~\cite{nagamatsu5d},
Buckminsterfullerenes~\cite{cohen5e},
Carbon-nanotubes~\cite{sheng5f} and unbelievably even in
DNA~\cite{kasumov5g}. It seems that superconductivity could be one
of the fundamental properties of nature that has been elusive
prior to Bednorz's discovery. Cuprates and its potential
applications were also extensively publicized to the extent where
it initiated the superconductivity-renaissance, which was between
late 80s and early 90s due to exceptional increase of $T_c$ from
about 20 K to 90 K within months. In those days before 1986,
reports on superconductivity were revolved around simple metals,
metallic alloys and some organic salts. Perhaps due to bad omen,
say brought about by the {\it black} schrodinger cat, scientists
have had a hard time to find the correct mechanism for
conventional superconductors, which nearly took them 50 years to
arrive at BCS theory in 1957. Hopefully history will not repeat
itself for the superconducting cuprates that further perplex
scientists with its puzzling phenomena. Cuprates, the unique oxide
has many {\it faces} depending upon temperature and doping unlike
other known superconductors and electronic materials reported thus
far. One of the {\it face} is of course, superconducting region
but there are more than meets the eye such as antiferromagnetic
fluctuations, pseudogap, Fermi liquid and non-Fermi liquid as one
dopes the cuprates or varies the temperature. As a matter of fact,
there have been sufficient amount of experimental results on
cuprates that could somehow pin-point the proper ingredients for
at least to understand the normal state transport properties of
high-$T_c$ superconductors to a certain reasonable extent.
Anisotropic phenomena on the normal state resistivity ($\rho(T)$)
curves such as the existence of metallic-like conduction,
semiconducting property below a certain crossover-temperature
($T_{crossover}$) and incompatibility of $T_{crossover}$ in
$c$-axis with characteristic temperature ($T^*$) in
$ab$-plane~\cite{arulsamy6,arulsamy7} point towards that
$T_{crossover}$ in $\rho_c(T)$ could be a different entity from
$T^*$ in $\rho_{ab}(T)$. Therefore, it is possible to work-out the
details of transport properties in $c$-axis in accordance with
ionization energy based Fermi-Dirac statistics (iFDS) in which the
charge carriers is further confined with an additional parameter
known as the ionization energy ($E_I$). In the procedure of
deriving iFDS, a somewhat different distribution functions for
holes and electrons are also obtained that further justifies the
inappropriateness of iFDS for systems consist of free electrons
and also for systems which are influenced by strong
electron-phonon scattering or both~\cite{arulsamy8,arulsamy8a}. It
is a well established fact that the transport dimensionality in
the normal state of high-$T_c$ superconductors is 2-dimensional
(2D). As such, comprehensive derivation of iFDS will be
carried-out to model the 2D equations of charge carriers'
concentrations ($\sqrt{np}(T,E_I)$ and $n(T,E_I)$) and
resistivities ($\rho_{poly}(T,E_I)$ and $\rho_c(T,E_I)$) as a
function of temperature ($T$) in the vicinity of electric field
that will eventually enable one to enumerate $\sqrt{np}(T,E_I)$,
$n(T,E_I)$, $\rho_{poly}(T,E_I)$ and $\rho_c(T,E_I)$. Gapless
nature of charge carriers in $ab$-planes as a consequence of the
experimental results by Basov {\it et al.}~\cite{basov9} is due to
an interesting discrepancy between $ab$-plane and $c$-axis in term
of superfluid density ($\rho_s$) and spectral weights in the
normal ($N_n$) and superconducting states ($N_s$). In $c$-axis,
$\rho_s$ $>$ $N_n$ $-$ $N_s$ while $\rho_s$ $=$ $N_n$ $-$ $N_s$ in
$ab$-plane, which signifies a gapless behavior of charge carriers
in $ab$-planes~\cite{arulsamy10}. This result will also be
incorporated into $\rho_{poly}(T,E_I)$ to obtain $\rho_{ab}(T)$.
Previously, Anderson and Zou~\cite{anderson11} proposed an
empirical resistivity equation given by

\begin {eqnarray}
\rho & = & \frac{A}{T} + BT,\label{eq:1}
\end {eqnarray}

for both $ab$-plane and $c$-axis. Equation~(\ref{eq:1}) simply suggests that there are two types of competing $T$-dependences in the normal state of high-$T_c$ superconductors. It is quite obvious that the coexistence of metallic and semiconducting resistivities in both $\rho_{ab}(T)$ and $\rho_c(T)$ are intriguing. Basically, $\rho_c(T,E_I)$ transition occurring from metallic (normal state) superconductor (MS) to metallic-semiconducting crossover (normal state; without $T^*$) superconductor (MSS) or vice versa with doping is shown to be predictable with models derived from iFDS. It is to be noted that the significance of $T^*$ on $\rho_{ab}(T)$ is ignored in this work since it is still controversial both in terms of experimental and theoretical approaches~\cite{arulsamy6,arulsamy7}. Parallel to this, $\rho_{ab}(T)$ and $R_H^{(ab)}(T)$ involving $T^*$ are yet another critical sub-aspects of the above-mentioned conduction peculiarities that need further studies. A theoretical proposal on these phenomena that consist of electrons, spinons and holons was given in Refs.~\cite{arulsamy6,arulsamy7} and references therein. Finally, superconducting samples of both crystalline and polycrystalline reported by various research groups will be employed extensively to accentuate the importance of iFDS in the normal state of 2D high-$T_c$ superconductors.

\section{Ionization energy based Fermi-Dirac statistics}\lb{s-eqs}

The conduction electron's distribution can be derived using
iFDS with ionization energy as an anomalous
constraint. This derivation involves two restrictive conditions:
(i) the total number of electrons in a given system is constant
and (ii) the total energy of $n$ electrons in that system is also
constant. Both conditions are as given below

\begin {eqnarray}
\sum_i^{\infty} n_i & = & n,\label{eq:2}
\end {eqnarray}

\begin {eqnarray}
\sum_i^{\infty} E_i n_i & = & E.\label{eq:3}
\end {eqnarray}

However, the condition as given in Eq.~(\ref{eq:3}) must be rewritten as given in
Eq.~(\ref{eq:4}) by
inserting conditions, $E_{electron}$ = $E_{initial state}$ + $E_I$
and $E_{hole}$ = $E_{initial state}$ $-$ $E_I$ appropriately.

\begin {eqnarray}
\sum_i^{\infty} (E_{initial state}\pm E_I)_i n_i & = & E.\label{eq:4}
\end {eqnarray}

This is to justify that an electron to occupy a higher state $N$ from
initial state $M$ is more probable than from initial state $L$ if
condition $E_I(M)$ $<$ $E_I(L)$ at certain $T$ is
satisfied. As for a hole to occupy a lower state $M$ from initial
state $N$ is more probable than to occupy state $L$ if the same
condition above is satisfied. $E_{initial state}$ ($E_{is}$) is the energy of
a particle in a given system at a certain initial state and ranges
from $+\infty$ to 0 for electrons and 0 to $-\infty$ for holes.
The importance of this inclusion is that it can
be interpreted as a gap that will be described later and also,
particularly the $E_I$ can be used to estimate the resistivity transition upon substitution
of different valence state ions. Utilizing Eq.~(\ref{eq:4}), one can write the mean number of particles in a certain quantum state $s$ ($\overline{n}_s$) as

\begin{eqnarray}
\overline{n}_s =
\frac{\sum\limits_{n_1,n_2,...} n_s \exp [-\lambda [n_1(E_{is} \pm E_I)_1 + n_2(E_{is} \pm E_I)_2 +...+ n_s(E_{is} \pm E_I)_s +...]]} {\sum\limits_{n_1,n_2,...} \exp [-\lambda [n_1(E_{is} \pm E_I)_1 + n_2(E_{is} \pm E_I)_2 +...+ n_s(E_{is} \pm E_I)_s +...]]}.
\label{eq:5}
\end{eqnarray}

Since state $s$ was chosen arbitrarily to calculate the mean number of particles in it thus Eq.~(\ref{eq:5}) has to be algebraically rearranged as given below to ease the following derivation

\begin{eqnarray}
\overline{n}_s = \frac {\sum\limits_{n_s} n_s \exp[-\lambda n_s(E_{is} \pm E_I)_s]\sum\limits_{n_1,n_2,...}^{(s)} \exp [-\lambda [n_1(E_{is} \pm E_I)_1 + n_2(E_{is} \pm E_I)_2 +...]]}{\sum\limits_{n_s} \exp[-\lambda n_s(E_{is} \pm E_I)_s]\sum\limits_{n_1,n_2,...}^{(s)} \exp [-\lambda [n_1(E_{is} \pm E_I)_1 + n_2(E_{is} \pm E_I)_2 +...]]}.
\label{eq:6}
\end{eqnarray}

Note that the sum, $\sum_{n_1,n_2,...}^{(s)}$ omits the chosen state, $s$. Subsequently, the partition function that represents $\sum_{n_1,n_2,...}^{(s)}$ can be written as

\begin{eqnarray}
Z_s(n) = \sum\limits_{n_1,n_2,...}^{(s)}\exp[-\lambda [n_1(E_{is} \pm E_I)_1 + n_2(E_{is} \pm E_I)_2 +...]].
\label{eq:7}
\end{eqnarray}

Fermi-Dirac statistics, including iFDS also need to satisfy Pauli's exclusion principle apart from the restrictive conditions given in Eqs.~(\ref{eq:2}) and~(\ref{eq:4}) in which, Pauli's principle requires that one needs to sum over $n_s$ = 0 and 1. This implies that each state can only accomodate one particle. In doing so, one arrives at Eq.~(\ref{eq:8}) from Eqs.~(\ref{eq:6}) and~(\ref{eq:7}).

\begin{eqnarray}
\overline{n}_s  &&= \frac {0 + \exp [-\lambda n_s(E_{is} \pm E_I)_s]Z_s(n-1)}{Z_s(n) + \exp [-\lambda n_s(E_{is} \pm E_I)_s]Z_s(n-1)}\nonumber \\&&
= \frac {1}{\frac {Z_s(n)}{Z_s(n-1)}\exp [\lambda n_s(E_{is} \pm E_I)_s] + 1}.
\label{eq:8}
\end{eqnarray}

Further simplification of Eq.~(\ref{eq:8}) can be performed using the relation (assuming $n$ $\gg$ 1)

\begin{eqnarray}
&&\ln Z_s(n-1) = \ln Z_s(n) - \frac {\partial \ln Z_s}{\partial n}, \nonumber \\&&
Z_s(n-1) = Z_s(n) \exp \left[-\frac {\partial \ln Z_s}{\partial n}\right].
\label{eq:9}
\end{eqnarray}

Taking $\partial \ln Z_s$/$\partial n$ = $\mu_s$ = $\mu$ as a consequence of large $n$ and substituting it into Eq.~(\ref{eq:8}) will lead one to write

\begin{eqnarray}
\overline{n}_s = \frac {1}{\exp [\mu + \lambda(E_{is} \pm E_I)_s] + 1}.
\label{eq:10}
\end{eqnarray}

Equation~(\ref{eq:10}) is the ionization energy based Fermi-Dirac statistics and the procedure for deriving it, is identical to standard Fermi-Dirac statistics that is given in Ref.~\cite{reif15}.

\section{Resistivity models}\lb{s-eqs}

The probability functions for electrons and holes are derived from Eq.~(\ref{eq:10}) after further considering $\exp[\mu + \lambda(E \pm E_I)]$ $\gg$ 1 which are given in Eqs.~(\ref{eq:11}) and ~(\ref{eq:12}) respectively~\cite{arulsamy10}

\begin{eqnarray}
f_e(E,E_I) = \exp \big[-\mu - \lambda(E + E_I)\big],
\label{eq:11}
\end{eqnarray}

\begin{eqnarray}
f_h(E,E_I) = \exp\big[\mu + \lambda(E - E_I)\big].
\label{eq:12}
\end{eqnarray}

The parameters $\mu$ and $\lambda$ are the normalization constants that will be derived shortly from Eqs.~(\ref{eq:2}) and~(\ref{eq:4}) respectively. In the standard FDS,
the respective Eqs.~(\ref{eq:11}) and~(\ref{eq:12}) are simply given by

\begin{eqnarray}
f_e(E) = \exp \big[-\mu - \lambda E \big],
\label{eq:13}
\end{eqnarray}

\begin{eqnarray}
f_h(E) = \exp\big[\mu + \lambda E \big].
\label{eq:14}
\end{eqnarray}

The 2D density of states' (DOS) derivative can be shown to be

\begin{eqnarray}
\frac{dN}{d{\bf k}} = \frac{L^2{\bf k}}{2\pi}.\label{eq:15}
\end{eqnarray}

Subsequently, the restrictive condition as given in Eq.~(\ref{eq:2}) can be rewritten for both electrons and holes respectively by employing the above-stated 2D $dN$/$d${\bf k} as

\begin{eqnarray}
n & = &\frac {L^2}{2\pi}e^{-\mu - \lambda E_I} \int\limits_0^\infty {\bf k} \exp\left(-\lambda
\frac{\hbar^2{\bf k}^2}{2m}\right) d{\bf k},\label{eq:16}
\end{eqnarray}

\begin{eqnarray}
p & = & \frac {L^2}{2\pi}e^{\mu - \lambda E_I} \int\limits_{-\infty}^0 {\bf k} \exp\left(\lambda
\frac {\hbar^2{\bf k}^2}{2m}\right) d{\bf k}.\label{eq:17}
\end{eqnarray}

Note here that $E$ is substituted with $\hbar^2{\bf k}^2$/$2m$. $m$ is the mass of the charge carriers while $n$ and $p$ are the respective concentrations of electrons and holes.
$L^2$ denotes area in
{\bf k}-space and $\hbar$ = $h/2\pi$, $h$ is the Planck constant. The respective solutions of Eqs.~(\ref{eq:16}) and~(\ref{eq:17}) are given below

\begin{eqnarray}
e^{\mu + \lambda E_I} & = & \frac{mL^2}{2n\pi\lambda\hbar^2},\label{eq:18}
\end{eqnarray}

\begin{eqnarray}
e^{\mu - \lambda E_I} & = & \frac{2p\pi\lambda\hbar^2}{mL^2},\label{eq:19}
\end{eqnarray}

The above-stated solutions were obtained via integral

\begin{eqnarray}
\int\limits_0^{\infty} {\bf x}^{2n+1} \exp(-a{\bf x}^2) d{\bf x} = \frac{n!}{2a^{n+1}}.\nn
\end{eqnarray}

Equations~(\ref{eq:18}) and~(\ref{eq:19}) respectively imply that~\cite{arulsamy8a}

\begin{eqnarray}
\mu_e(iFDS) = \mu_e + \lambda E_I, \label{eq:20}
\end{eqnarray}

\begin{eqnarray}
\mu_h(iFDS) = \mu_h - \lambda E_I. \label{eq:21}
\end{eqnarray}

On the other hand, using Eq.~(\ref{eq:4}), one can obtain

\begin{eqnarray}
E && = \frac {L^2\hbar^2}{4m\pi}  e^{-\mu -\lambda E_I} \int\limits_0^\infty {\bf k}^3
\exp\left(\frac{-\lambda\hbar^2{\bf k}^2}{2m}\right)d{\bf k} \nn \\&& =
\frac{m}{2\pi}\bigg(\frac{L}{\lambda\hbar}\bigg)^2 e^{-\mu -\lambda E_I}.
\label{eq:22}
\end{eqnarray}

Equation~(\ref{eq:22}) after appropriate substitution with Eq.~(\ref{eq:18}) is compared with the energy of a 2D ideal gas given below

\begin{eqnarray}
E = nk_BT. \label{eq:23}
\end{eqnarray}

Quantitative comparison will eventually leads to $\lambda_{iFDS}$ = $\lambda_{FDS}$ = 1/$k_BT$ where $k_B$ is the Boltzmann constant. Actually, equation~(\ref{eq:23}) can be derived trivially with the procedure given in Refs.~\cite{resnick16,reif15}. Identical $\lambda$ between FDS and iFDS has been somewhat anticipated and is not unusual either. Nevertheless, Eqs.~(\ref{eq:20}) and~(\ref{eq:21}) will have enormous consequences on the properties of charge carriers as will be pointed out in the following paragraphs.

The distribution function for electrons and holes can be written explicitly by first denoting $\mu$ $=$ $-E_F$ (Fermi level), $\lambda$ $=$ 1/$k_BT$ and substituting these into Eqs.~(\ref{eq:11}) and~(\ref{eq:12}) will lead one to write

\begin{eqnarray}
f_e(E,E_I) &=& \exp\left[\frac{E_F - E_I - E}{k_BT}\right]. \label{eq:24}
\end{eqnarray}

\begin{eqnarray}
f_h(E,E_I) &=& \exp\left[\frac{E - E_I - E_F}{k_BT}\right]. \label{eq:25}
\end{eqnarray}

It is worth noting that, $-E_I$ in Eq.~(\ref{eq:25}) for holes is a natural consequence of Dirac's theory of antiparticle interpretations~\cite{sakurai17}. Physically, $E_I$ is due to electron-ion attraction or Coulombic in nature that has to be contrasted with band gap, $E_g$ which arises from energy band splitting. General equations to compute charge carriers' concentrations are as listed below

\begin{eqnarray}
n &=& \int\limits^{\infty}_0{f_e(E,E_I)N_e(E)dE}.
\label{eq:26}
\end{eqnarray}

\begin{eqnarray}
p &=& \int\limits_{-\infty}^0{f_h(E,E_I)N_h(E)dE}.
\label{eq:27}
\end{eqnarray}

$N_e(E)$ and $N_h(E)$ are the 2D DOS for electrons and holes respectively. It is now convenient to obtain
the geometric-mean concentrations of electrons and holes in the normal state of 2D polycrystalline high-$T_c$ superconductors as~\cite{arulsamy10}

\begin{eqnarray}
\sqrt{np}(T,E_I) & = &
\frac{k_BT}{\pi\hbar^2}\big(m_e^*m_h^*\big)^{1/2}\exp\left[\frac{-E_I}{k_BT}\right],
\label{eq:28}
\end{eqnarray}

Here, $n$ $\approx$ $p$ is assumed to avoid $E_F$ and the 2D DOS is given by

\begin{eqnarray}
N_{e,h}(2D) = \frac{m^*_{e,h}}{\pi\hbar^2}.
\label{eq:29}
\end{eqnarray}

Note here that $N_{e,h}$(2D) is independent of $E$ and $m^*_{e,h}$ is the effective mass of either electrons or holes. In order to calculate $\rho_c(T,E_I)$, $n$ needs to be determined first that is given by (from Eq.~(\ref{eq:26}))

\begin{eqnarray}
n(T,E_I) & = &
\frac{m_e^*k_BT}{\pi\hbar^2}\exp\left[\frac{E_F - E_I}{k_BT}\right].
\label{eq:30}
\end{eqnarray}

The 2D resistivity models for polycrystals and in $c$-axis can be derived from an elementary resistivity equation,

\begin{eqnarray}
\rho = \frac{m^*}{ne^2\tau}.
\label{eq:31}
\end{eqnarray}

Substituting, 1/$\tau$ $=$ $AT^2$ as a consequence of electron-electron interactions and Eq.~(\ref{eq:28}) for $n$ into Eq.~(\ref{eq:31}) gives~\cite{arulsamy10}

\begin{eqnarray}
\rho_{poly}(T,E_I) & = &
\frac{A_2\pi\hbar^2}{e^2k_B}T \exp\left[\frac{E_I}{k_BT}\right],
\label{eq:32}
\end{eqnarray}

Similarly, $\rho_c(T,E_I)$ can be shown to be~\cite{arulsamy8}

\begin{eqnarray}
\rho_c(T,E_I) & = &
\frac{A_2\pi\hbar^2}{e^2k_B}T \exp\left[\frac{E_I - E_F}{k_BT}\right],
\label{eq:33}
\end{eqnarray}

Note that $A_2$ is a 2D $T$-independent scattering rate constant in the absence of magnetic field and $e$ is the charge of an electron. Furthermore, one can recall the characteristics of Eq.~(\ref{eq:1}) which has been captured by Eqs.~(\ref{eq:32}) and ~(\ref{eq:33}) where both equations also contain two competing $T$-dependences though in a different proportionality. One can also surmise that for 2D systems, $k_BT_{crossover}$ = $E_I$ for $\rho_{poly}(T)$ and $k_BT_{crossover}$ = $E_I - E_F$ for $\rho_c(T)$. Now, applying the result of gapless nature of charge carriers in $ab$-planes into Eq.~(\ref{eq:32}) leaves one with~\cite{arulsamy10}

\begin{eqnarray}
\rho_{ab}(T) & = &
\frac{A_2\pi\hbar^2}{e^2k_B}T.
\label{eq:34}
\end{eqnarray}

The nature of gapless phenomena in $ab$-planes could be due to change in scattering rate that originates from a different Fermi phase space argument in accordance with nested Fermi liquid theory~\cite{virosztek18}. Note that Eq.~(\ref{eq:34}) is nothing but a special case that satisfy, $E_I$ $\ll$ $T_c$ inequality. It is stressed that the influence of free electrons and $T$-dependence of electron-phonon scattering were entirely ignored in the $\rho_{poly}(T,E_I)$ and $\rho_c(T,E_I)$ models as given in Eqs.~(\ref{eq:32}) and~(\ref{eq:33}) respectively. Hence, those models are obviously not suitable for conventional metals. Interestingly, the electrons of a 2D semiconductor with $E_g$ can be further gapped with $E_I$ that will lead one to derive $\sqrt{np}(T,E_I,E_g)$ and $\rho(T,E_I,E_g)$ respectively as

\begin{eqnarray}
\sqrt{np}(T,E_I,E_g) =
\frac{k_BT}{\pi\hbar^2}\big(m_e^*m_h^*\big)^{1/2}\exp\left[\frac{-E_I - \frac{1}{2}E_g}{k_BT}\right].\label{eq:35}
\end{eqnarray}

\begin{eqnarray}
\rho(T,E_I,E_g) & = &
\frac{A_2\pi\hbar^2}{e^2k_B}T \exp\left[\frac{E_I + \frac{1}{2}E_g}{k_BT}\right],
\label{eq:36}
\end{eqnarray}

Consequently, electrons in the conduction band ($E$ $>$ $E_g$) will still be influenced by $E_I$ even though $E_I$ $<$ $E_g$ due to $E_I$'s Coulombic nature between charge carriers and ions.

\section{Discussion}\lb{s-eqs}

The effect of Nd$^{3+}$
($E_I$ = 1234 kJmol$^{-1}$) substitution into Sr$^{2+}$ ($E_I$ = 807 kJmol$^{-1}$)
in superconducting TlSr$_{2-x}$Nd$_x$CaCu$_2$O$_7$ compound~\cite{shukor21} was found
to increase the $\rho(T)$ in accordance with $E_I$. This justifies the need for $E_I$ based
analysis on doping as pointed out by iFDS. Applications of iFDS in superconductors are
explicitly given in Refs.~\cite{arulsamy6,arulsamy7,arulsamy8,arulsamy10}. Recently,
Naqib {\it et al}.~\cite{naqib22} have investigated the electrical properties of
Y$_{1-x}$Ca$_x$Ba$_2$(Cu$_{1-y}$Zn$_y)_3$O$_{7-d}$ superconducting compounds by varying
$x$, $y$ and $d$. The transition of normal state $\rho(T)$ with Ca$^{2+}$
($E_I$ = 867 kJmol$^{-1}$) doped into Y$^{3+}$ ($E_I$ = 1253 kJmol$^{-1}$) sites is in excellent
agreement with Eq.~(\ref{eq:32}) of iFDS. But,
Zn$^{2+}$ doping is not appropriate to analyze as a function of iFDS only
because this substitution will directly disturb the $ab$-plane conduction
of spinons and holons and also in term of oxygen concentration ($d$), thus
the overall conductivity of Y$_{1-x}$Ca$_x$Ba$_2$(Cu$_{1-y}$Zn$_y)_3$O$_{7-d}$
polycrystals will be modified in a not-so-simple way~\cite{arulsamy6,arulsamy7}.
It is easy however, to extract the
relation of normal state $\rho(T)$ between Y$_{0.9}$Ca$_{0.1}$Ba$_2$Cu$_3$O$_{7-d}$ and
Y$_{0.8}$Ca$_{0.2}$Ba$_2$Cu$_3$O$_{7-d}$ where the normal state $\rho(T)$
is reduced with Ca$^{2+}$ doping for all $d$ (oxygen pressure),
since Y$^{3+}$($E_I$ $=$ 1253 kJmol$^{-1}$) $>$ Ca$^{2+}$($E_I$ $=$ 867 kJmol$^{-1}$).

In contrast, Sr$^{2+}$ ($E_I$ = 807 kJmol$^{-1}$) substitution
into Ba$^{2+}$ ($E_I$ = 734 kJmol$^{-1}$) sites have decreased the
normal state $\rho(T)$ in
Hg$_{0.85}$Re$_{0.15}$(Ba$_{1-y}$Sr$_y$)$_2$Ca$_2$Cu$_3$O$_{8-\delta}$
unexpectedly~\cite{batista23}. iFDS suggests that $\rho(T)$ should
increase with Sr doping into Ba sites since Sr$^{2+}$ ($E_I$ $=$
807 kJmol$^{-1}$) $>$ Ba$^{2+}$ ($E_I$ $=$ 734 kJmol$^{-1}$). This
contradicting scenario can be explained since the actual doping
concentration determined with EDX showed that the concentrations
of other elements namely, Re, Ca and Cu also vary with Sr doping
into Ba sites. It is quite non-trivial to verify and prove these
by calculating the relative $E_I$ for Ca$^{2+}$, Sr$^{2+}$,
Ba$^{2+}$, Re and Cu$^{2+,3+}$. For comparison purposes, the
valence state of Re and Cu are taken to be 3+ and 2+ respectively.
Arbitrary values of valence state for Re and Cu will {\it not}
affect this analysis since its valence states are assumed to be
constant (in this case) for Sr00, Sr17 and Sr28. It is known from
Ref.~\cite{batista23} that the concentrations and $E_I$ for Sr00
are Re$^{3+}$ (0.15; 1510 kJmol$^{-1}$), Ba$^{2+}$ (2.10; 734
kJmol$^{-1}$), Sr$^{2+}$ (0.00; 807 kJmol$^{-1}$), Ca$^{2+}$
(2.20; 867 kJmol$^{-1}$) and Cu$^{2+}$ (3.10; 1352 kJmol$^{-1}$).
For Sr17, it is given by Re$^{3+}$ (0.15; 1510 kJmol$^{-1}$),
Ba$^{2+}$ (0.84; 734 kJmol$^{-1}$), Sr$^{2+}$ (0.17; 807
kJmol$^{-1}$), Ca$^{2+}$ (1.97; 867 kJmol$^{-1}$) and Cu$^{2+}$
(3.12; 1352 kJmol$^{-1}$). Finally for Sr28, Re$^{3+}$ (0.14; 1510
kJmol$^{-1}$), Ba$^{2+}$ (0.74; 734 kJmol$^{-1}$), Sr$^{2+}$
(0.28; 807 kJmol$^{-1}$), Ca$^{2+}$ (1.75; 867 kJmol$^{-1}$) and
Cu$^{2+}$ (3.02; 1352 kJmol$^{-1}$). Therefore, from this data it is possible to
calculate the changes of $E_I$ due to the fluctuations of
other non-dopant elements' concentrations with Sr doping into Ba
sites. One can show that the relative $E_I$s are as given below
for Sr00, Sr17 and Sr28 respectively.

\begin{eqnarray}
E_I^{(Sr00)} &&= [0.15(1510)]{\bf _{Re^{3+}}} + [2.10(734)]{\bf _{Ba^{2+}}} \nn \\&&
~~+ [0.00(807)]{\bf _{Sr^{2+}}} + [2.20(867)]{\bf _{Ca^{2+}}} \nn \\&&
~~+ [3.10(1352)]{\bf _{Cu^{2+}}}\nn \\&&
= 7640~{\bf kJmol^{-1}},\label{eq:37}
\end{eqnarray}

\begin{eqnarray}
E_I^{(Sr17)} &&= [0.15(1510)]{\bf _{Re^{3+}}} + [0.84(734)]{\bf _{Ba^{2+}}} \nn \\&&
~~+ [0.17(807)]{\bf _{Sr^{2+}}} + [1.97(867)]{\bf _{Ca^{2+}}} \nn \\&&
~~+ [3.12(1352)]{\bf _{Cu^{2+}}} \nn \\&&
= 6680~{\bf kJmol^{-1}},\label{eq:38}
\end{eqnarray}

\begin{eqnarray}
E_I^{(Sr28)} &&= [0.14(1510)]{\bf _{Re^{3+}}} + [0.74(734)]{\bf _{Ba^{2+}}} \nn \\&&
~~+ [0.28(807)]{\bf _{Sr^{2+}}} + [1.75(867)]{\bf _{Ca^{2+}}} \nn \\&&
~~+ [3.02(1352)]{\bf _{Cu^{2+}}}  \nn \\&&
= 6369~{\bf kJmol^{-1}}.\label{eq:39}
\end{eqnarray}

Hence, the reduction of
$\rho(T)$ with Sr doping is justified from
Eqs.~(\ref{eq:37}),~(\ref{eq:38}) and~(\ref{eq:39}), which is due
to the concentration's fluctuation of Ca, Re and Cu apart from Ba
and Sr. All values of $E_I$ including in Eqs.~(\ref{eq:37}),~(\ref{eq:38})
and~(\ref{eq:39}) were averaged in accordance with

\begin{eqnarray}
E_I [X^{z+}] = \sum\limits_{i=1}^z\frac{E_{Ii}}{z},\label{eq:40}
\end{eqnarray}

and should not be taken literally since those $E_I$s
are not absolute values. The absolute values need to be obtained
from Eq.~(\ref{eq:41}) below~\cite{arulsamy8a}

\begin{eqnarray}
E_I = \frac{e^2}{8 \pi \epsilon \epsilon_0 r_B}.\label{eq:41}
\end{eqnarray}

If one of the substituting ion is multivalence, then the valence state of that particular ion can be approximated using Eq.~(\ref{eq:42}) below~\cite{arulsamy10}

\begin{eqnarray}
\frac{\delta}{j}\sum^{z+j}_{i=z+1}{E_{Ii}} + \frac{1}{z}\sum^{z}_{i=1}{E_{Ii}} & = & \frac{1}{q}\sum^{q}_{i=1}{E_{Ii}}.\label{eq:42}
\end{eqnarray}

The variations of $\rho(T)$ with doping in polycrystalline samples namely, (Pr$_{1-x}$Gd$_x$)$_{1.85}$Ce$_{0.15}$CuO$_4$ and (Pr$_{1-x}$Y$_x$)$_{1.85}$Ce$_{0.15}$CuO$_4$ from Meen {\it et al}.~\cite{meen24}, Pr$_x$Gd$_{1-x}$Ba$_2$Cu$_3$O$_{7-\delta}$ from Khosroabadi {\it et al}.~\cite{khosroabadi25} can be explained in a straight-forward manner without even using Eq.~(\ref{eq:42}). Meen {\it et al}.~\cite{meen24} found that substitutions of Gd$^{3+}$ and Y$^{3+}$ into Pr$^{3+}$ give rise to $\rho(T)$ simply because $E_I$(Pr$^{3+}$ = 1211 kJmol$^{-1}$) $<$ $E_I$(Gd$^{3+}$ = 1251 kJmol$^{-1}$) and $E_I$(Pr$^{3+}$) $<$ $E_I$(Y$^{3+}$ = 1253 kJmol$^{-1}$). Similarly, Khosroabadi {\it et al}.~\cite{khosroabadi25} observed decreasing $\rho(T)$ for Pr$^{3+}$ substitution into Gd$^{3+}$ due to former inequality. Further indirect observation was given by Isawa {\it et al}.~\cite{isawa26} for Ce$_x$Pr$_{1-x}$La$_2$CuO$_{4+\delta}$ polycrystals. However, the valence state of Ce is unknown in that sample therefore one has to employ Eq.~(\ref{eq:42}). The first term, $\frac{\delta}{j}\sum^{z+j}_{i=z+1}{E_{Ii}}$ in Eq.~(\ref{eq:42}) has $i$ $=$ $z$ + 1, $z$ + 2,..., $z$ + $j$ and $j$ $=$ 1, 2,
3,.... It is solely due to multivalence ion. In this case
Ce$^{3+,4+}$ is substituted into Pr$^{3+}$ sites
hence from Eq.~(\ref{eq:42}), the first term is due to Ce$^{4+}$ ion's
contribution or caused by reaction of the form Ce$^{3+}$ $-$
electron $\to$ Ce$^{4+}$ (3547 kJmol$^{-1}$), hence $j$ is equals to 1 here
and $\delta$ represents the additional contribution from
Ce$^{4+}$. The second ($i$ $=$ 1, 2, 3, ..., $z$) and last ($i$
$=$ 1, 2, 3, ..., $q$) terms respectively are due to reaction of
the form Ce $-$ 3(electrons) $\to$ Ce$^{3+}$ and Pr $-$
3(electrons) $\to$ Pr$^{3+}$. Recall that $q$ = $z$ = 3+ and $i$ =
1, 2, 3,... represent the first, second, third, ... ionization
energies while $j$ = 1, 2, 3, ... represent the fourth, fifth,
sixth, ... ionization energies. Therefore, $z$ + $\delta$ gives
the minimum valence number for Ce which can be calculated from
Eq.~(\ref{eq:42}). Apparently, Ce$^{3+}$ ($E_I$ = 1178 kJmol$^{-1}$) substitution into Pr$^{3+}$ ($E_I$ = 1211 kJmol$^{-1}$) will lead to a lower $\rho(T)$ as indicated in Ref.~\cite{isawa26}. The opposite scenario, increment of $\rho(T)$ with Ce$^{3+\delta}$ is possible if and only if the valence state of Ce is $>$ 3.010. This value (3.010) is actually obtained from Eq.~(\ref{eq:42}).

Interestingly, Shi {\it et al}.~\cite{shi27} have reported substitution effects of Ru$_{1-x}$Sb$_x$Sr$_2$Gd$_{1.4}$Ce$_{0.6}$Cu$_2$O$_{10-\delta}$ polycrystals on $\rho(T)$. Since $E_I$(Sb$^{3+}$ = 1623 kJmol$^{-1}$) $<$ $E_I$(Ru$^{3+}$ = 1692 kJmol$^{-1}$) then it is quite obvious to predict $\rho(T)$ will be reduced in magnitude with Sb$^{3+}$ content. However, a contradicting scenario have been observed where $\rho(T)$ increases with Sb. As such, Eq.~(\ref{eq:42}) is again useful to estimate the minimum valence state of Sb$^{3+\delta}$ in order to justify $\rho(T)$ increment with Sb. The minimum valence state of Sb has to be $>$ 3+ ($z$ + $\delta$) so as to satisfy $\rho(T)$ increment. $z$ + $\delta$ is calculated to be 3.017 or Sb$^{3.017+}$. Another point worth extracting is that additional annealing (4 days) for RuSr$_2$Gd$_{1.4}$Ce$_{0.6}$Cu$_2$O$_{10-\delta}$ (sample B) does not change $E_I$ significantly whereas the $A_2$ parameter is reduced dramatically compared to as-prepared sample (A). The assumption that $E_I$ changes insignificantly with annealing is valid since the shape of $\rho(T)$ curves for samples A and B are identical~\cite{shi27}. This phenomenon clearly indicate that long hours of proper heat-treatment improves sample B's quality in terms of grain boundaries, defects~\cite{shi27} and subsequently reduces the magnitude of scattering rates ($A_2$). Figure~\ref{fig1} indicates calculated plots of $\rho(T,E_I)$ for Hg$_{0.7}$Cr$_{0.3}$Sr$_2$CuO$_{4+\delta}$ from Ref.~\cite{kandyel28} and Gd$_{0.95}$Pr$_{0.05}$Ba$_2$Cu$_3$O$_{7-\delta}$ from Ref.~\cite{khosroabadi25} while the inset is for Sr$_{0.9}$La$_{0.1}$CuO$_2$ from Ref.~\cite{karimoto29} and TmBa$_2$Cu$_3$O$_{6.99}$ from Ref.~\cite{sulkowski30}. Note that the symbols do not represent the experimental data points but simply to enhance contrast between all those calculated curves of different compositions. $\rho(T)$ curves were obtained by fitting experimental data with Eq.~(\ref{eq:32}). Table 1 lists all the fitting parameters namely, $T$-independent scattering rate constant ($A_2$), charge carriers concentrations ($\sqrt{np}$), charge gap parameter ($E_I$) and d$\rho(T)$/d$T$ in detail above $T_c$ and $T^*$. $\sqrt{np}$ is estimated by first computing $E_I$ from Eq.~(\ref{eq:32}) and then substitute it into Eq.~(\ref{eq:28}). Furthermore, two values of $\sqrt{np}$ are computed at 300 K, in which one satisfies $m^*/m_o$ = 1 while the other takes $m^*/m_o$ = 50. $m_o$ is the rest mass of an electron and the fitted values of $A_2$ are  actually equal to $A_2\pi\hbar^2$/$e^2k_B$. Sulkowski {\it et al}.~\cite{sulkowski30}, have computed charge carriers concentrations from Hall effect measurements for TmBa$_2$Cu$_3$O$_{6.99}$ that is in the order of 10$^{22}$ cm$^{-3}$. This value is comparable with the estimated value from Eq.~(\ref{eq:28}) with $m^*/m_o$ = 50, which is 4.7 $\times$ 10$^{18}$ m$^{-2}$ = 4.7 $\times$ 10$^{22}$ cm$^{-3}$. Parallel to this, Eqs.~(\ref{eq:32}) and~(\ref{eq:28}) also enable one to approximate the effective mass of charge carriers. As for single crystals, one may still employ Eq.~(\ref{eq:32}) via the assumption~\cite{arulsamy10}

\begin{eqnarray}
\rho_{poly}(T) &=& \sqrt{\rho_c(T)\rho_{ab}(T)}. \label{eq:43}
\end{eqnarray}

Eq.~(\ref{eq:43}) is assumed to be valid to convert both $\rho_c(T)$ and $\rho_{ab}(T)$ data to $\rho_{poly}(T)$, which will enable one to analyze the effect of $A_2$ parameter between single crystals and polycrystals. One should not substitute Eqs.~(\ref{eq:33}) and~(\ref{eq:34}) into Eq.~(\ref{eq:43}) because Eq.~(\ref{eq:34}) was derived based on a special case as mentioned previously. Subsequently, the changes on scattering rate ($A_2$) will be captured. Recently, Lanzara {\it et al}.~\cite{lanzara31} have shown quite convincingly via ARPES
measurements that the existence of electron-phonon coupling associated
with movements of oxygen atoms in Bi$_2$Sr$_2$CaCu$_2$O$_8$,
Bi$_2$Sr$_2$Cu$_2$O$_6$ and La$_{2-x}$Sr$_x$CuO$_4$ should not be
neglected entirely. This observation could be due to polarons that is well
represented by $E_I$ in iFDS. The important difference between polarons and free
electron-phonon scattering is that the latter has a very strong
$T$-dependence while the former increases the effective mass of
the charge carriers to some extent. This could be the sole reason
for the missing electron-phonon coupling effect on $\rho(T)$
measurements in high-$T_c$ superconducting cuprates thus far. All
$E_I$ values were obtained from Ref.~\cite{web32} and the
predictions stated above are only valid for reasonably pure
materials without any significant impurity phases and grain boundary effects.

\section{Conclusions}\lb{s-eqs}

In summary, iFDS is shown to be an appropriate theory in the normal state of high-$T_c$ superconductors. The variations on polycrystalline resistivity with doping were very well explained with one exception, in which the importance of characteristics temperature was ignored throughout this work partly due to vague understanding on temperature dependence of $c$-axis and $ab$-plane scattering rates. It is interesting to note that iFDS is able to predict the resistivity's magnitude with doping via ionization energy as well as enable one to estimate the valence state of substituting ions accurately. Apart from that, the fitting parameters give a reasonable prediction of the charge carriers' effective mass in the vicinity of 10$^{-29}$ kg or 50 times heavier than electron's rest mass. The concentration of this heavy charge carriers is around 10$^{22}$ cm$^{-3}$ whereas light charge carriers' concentration is approximately 10$^{20}$ cm$^{-3}$  when the effective mass is equals to rest mass. The heavier effective mass (1 $<$ $m^*/m_o$ $\le$ 50) somewhat imply the existence of polarons in the normal state of high-$T_c$ superconductors that could mask electron-phonon coupling effect on resistivity versus temperature measurements.

\section*{Acknowledgments}
ADA would like to thank the National University of Singapore for the financial aid and also to Prof. Feng Yuan Ping for his support. ADA personally thanks Saleh H. Naqib and his group for their kind approval of using their experimental data on cuprates prior to publication as well as Dr Igor Yurin for his helpful comments on Cooperon in Cuprates. The author is grateful and beholden to Arulsamy Innasimuthu, Sebastiammal Innasimuthu, Arokia Das Anthony and Cecily Arokiam, whom are the financiers of private (non-governmental) and not-for-profit Condensed Matter Group (CMG-Ampang), which was founded solely for the advancement of science.

\begin{figure}
\caption{Computed $\rho(T,E_I)$ curves for Hg$_{0.7}$Cr$_{0.3}$Sr$_2$CuO$_{4+\delta}$  and Gd$_{0.95}$Pr$_{0.05}$Ba$_2$Cu$_3$O$_{7-\delta}$ from Refs.~\cite{kandyel28} and~\cite{khosroabadi25} respectively. The inset is for Sr$_{0.9}$La$_{0.1}$CuO$_2$  and TmBa$_2$Cu$_3$O$_{6.99}$ from Refs.~\cite{karimoto29} and~\cite{sulkowski30} respectively. The curves were determined in accordance with Eq.~(\ref{eq:32}). The symbols were used solely to distinguish curves of different samples and it does not represent experimental data points. Apparently, $T_{crossover}$s for all these samples (if any) are $<$ $T_c$s, as such the metal to semiconductor transition is not observed, which has been anticipated because $E_I$ $<$ $k_BT_c$.}
\label{fig1}
\end{figure}

\begin{table}
\caption{Computed values of the temperature independent scattering rate constant ($A_2$), charge gap parameter ($E_I$), resistivity slope ($d\rho(T)/dT$) and the concentration of charge carriers ($\sqrt{np}$) at 300 K with different effective masses ($m^*/m_o$ = 1 and $m^*/m_o$ = 50) in the normal state of high-$T_c$ superconducting polycrystals. Note that the magnitudes of $E_I$ are in the range of 2-9 meV and the charge carriers' concentration is determined to be in the order of 10$^{18}$ m$^{-2}$ or 10$^{22}$ cm$^{-3}$ if $m^*/m_o$ = 50 is assumed. $d\rho(T)/dT$ is calculated from the experimental $\rho_{poly}(T)$ linear plots.}
\end{table}

\end{document}